\documentclass[aip,apl,reprint]{revtex4-1}
\usepackage{graphicx}
\usepackage{amsmath}

\begin{document}

\title{A three-dimensional Josephson parametric amplifier}

\author{Imran Mahboob}
\email{imran.mahboob@lab.ntt.co.jp}
\affiliation{NTT Basic Research Laboratories, NTT Corporation, Atsugi-shi, Kanagawa 243-0198, ~Japan}

\author{Hiraku Toida}
\affiliation{NTT Basic Research Laboratories, NTT Corporation, Atsugi-shi, Kanagawa 243-0198, ~Japan}

\author{Kousuke Kakuyanagi}
\affiliation{NTT Basic Research Laboratories, NTT Corporation, Atsugi-shi, Kanagawa 243-0198, ~Japan}

\author{Yasunobu Nakamura}
\affiliation{Research Center for Advanced Science and Technology, The University of Tokyo, Meguro-ku, Tokyo 153-8904, ~Japan}
\affiliation{Center for Emergent Matter Science, RIKEN, Wako-shi, Saitama 351-0198, ~Japan}

\author{Shiro Saito}
\affiliation{NTT Basic Research Laboratories, NTT Corporation, Atsugi-shi, Kanagawa 243-0198, ~Japan}

\begin{abstract}
{A Josephson parametric amplifier (JPA) is executed in a three-dimensional (3D) microwave cavity by coupling it to a superconducting quantum interference device (SQUID) that is embedded in a two-dimensional resonator. The JPA is activated in a three-wave mixing configuration by injecting ac magnetic flux, at twice the 3D cavity frequency, into the SQUID. An 8.3 GHz cavity is measured in a non-degenerate phase-insensitive configuration which yields gains in excess of 40 dB, where a 20 dB gain results in an operational bandwidth of 0.4 MHz, a 1 dB compression point of $-115$ dBm with half a quantum of added noise.}
\end{abstract}

\maketitle

The Josephson parametric amplifier, a superconducting resonator shunted with nonlinear inductance from Josephson junctions (JJs), was initially pioneered in the late 70s and early 80s to yield signal gain with low-noise and to study quantum optics with microwave flavoured photons \cite{jpa1, jpa2}. The JPA re-emerged in the late 2000s and it has become an indispensable component for quantum limited measurements at microwave frequencies \cite{qlm1, jpa3, jpa4, jpa5}. Indeed, JPAs have facilitated quantum non-demolition measurements of superconducting qubits \cite{jpaq1, jpaq2}, to monitor their relaxation and corresponding trajectories \cite{jpaq2}, to even apply feedback to this process \cite{jpaFB1, jpaFB2} as well as being instrumental in the precision measurements of mechanical vibrations, semiconductor quantum dots and surface acoustic waves \cite{jpaD1, jpaD2, jpaD3}. Moreover, the squeezed microwave radiation generated by the JPA has not only enabled quantum optics to be explored at ever greater extremes of non-classicality \cite{jpaQO1, jpaQO2, jpaQO3, jpaQO4, jpaQO5, jpaQO6} but it has also been employed to excite quantum states of a harmonic oscillator and to perform sub-vacuum noise measurements \cite{sqd1, sqd2}.

Meanwhile a concerted effort has taken place to develop a multi-superconducting-qubit architecture to enable the implementation of quantum error correction protocols which could ultimately lead to a fault-tolerant quantum computer \cite{qc1, qc2, qc3, qc4, qc5, qc6, qc7}. To that end, the ability to handle the large cumulative signal power from multiple qubits via a single JPA is essential for the realisation of this ambition. However, the second generation of JPAs from the late 2000s suffered from poor power handling capacity, whose gain became easily saturated, whilst operating over relatively narrow bandwidths. These shortcomings led to the emergence of third generation JPAs which sought to address these objectives simultaneously by either reducing the resonator's quality factor $Q$ \cite{jpa9, jpa10}, diluting the nonlinearity of the inductance with SQUID arrays \cite{jpa11, jpa12} or by eliminating the resonator entirely \cite{jpa13, jpa8}. Most recently, kinetic inductance parametric amplifiers have arisen which can operate without JJs and they have achieved unprecedented performance levels \cite{kipa1, kipa2}.

Here an alternative strategy is pursued, which is geared to simply increasing the dynamic range of the JPA, whilst neglecting the resultant bandwidth performance. To that end particular focus is paid to the source of gain saturation which stems from the anharmonicity in the restoring potential of the JPA and whose potency is a consequence of the direct incorporation of the nonlinear inductance into the amplifying resonator \cite{jpa11, jpa14}. By separating the participation of the nonlinear inductance in the signal amplifying resonator, by means of a dispersive interaction, its efficacy can be diluted.  On the other hand, the high $Q$ of the 3D cavity \cite{3Dp1} still enables tangible frequency modulations from the diluted nonlinear inductance thus enabling access to parametric amplification \cite{jpa3, jpa4}. For this 3D variant of a JPA, gains in excess of 40 dB are observed, which are large amongst the most competitive parametric amplification factors reported to date \cite{jpa5}.

\begin{figure}[!ht]
\vspace{-0.2cm}
\includegraphics[scale=1.0]{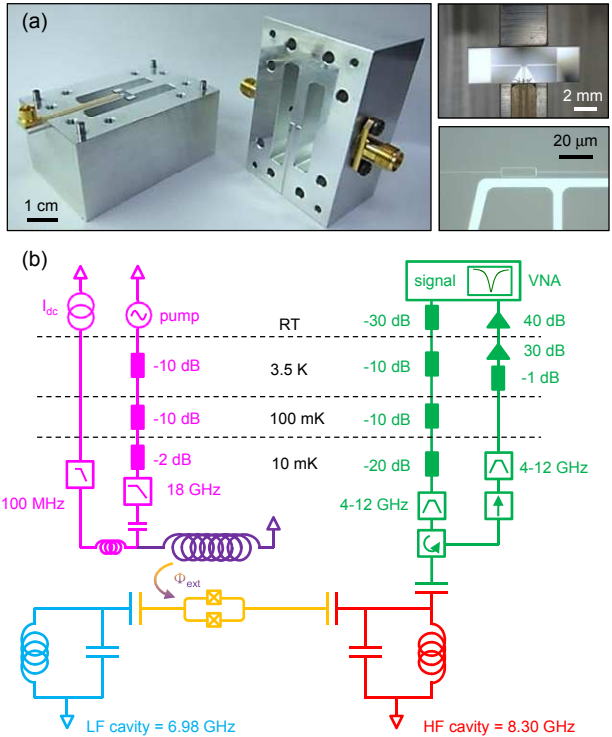}
\vspace{0.0cm}
\small{\caption{(a) A photograph of two 3D JPAs consisting of a pair of cavities that are milled from a split aluminium block (left) which enables access to the SQUID chip (top right) located in the centre where the SQUID (bottom right) is capacitively coupled to both cavities via the large rectangular antennas. The SQUID can be ac flux pumped and/or dc flux biased via application of current to the coplanar waveguide located between the cavities. TE$_{101}$ modes are hosted by the cavities that are probed via the SMA connectors to which external signals are coupled. (b) An equivalent circuit diagram of the 3D JPA and a schematic of the measurement circuit used to probe the cavities [HF (red) and LF (blue)]. The reflection coefficient $S_{11}$ is measured in a vector network analyser, VNA (green). The SQUID (yellow) has a nearby flux line (purple) to which external dc and ac current signals can be applied through a bias-tee (magenta).}}
\end{figure}

The 3D JPAs developed in this work are shown in Fig.~1(a) and they consist of two 3D cavities that are capacitively coupled to a magnetic flux pumped SQUID \cite{3DjpaV1}, whose equivalent electrical circuit is shown in the lower portion of Fig.~1(b). The smaller high-frequency (HF) cavity is detailed here and the larger lower-frequency (LF) cavity in Supplementary Materials \cite{SM}. It should be noted that both cavities can be operated as JPAs independently of each other. The HF cavity has a volume of $27 \times 5 \times 24$ mm$^3$ and is machined from aluminium (A1050) in a split configuration enabling access to the SQUID chip and a coplanar waveguide chip to inject magnetic flux into the former as shown in Fig.~1(a). A single SQUID is patterned via electron beam lithography on a high-resistance silicon substrate and metalized with double angle aluminium evaporation where the first aluminium layer is oxidised to form the JJs and the resultant chip is placed in a tunnel interconnecting the two cavities. The SQUID has a loop area of $20 \times 5$ $\mu$m$^2$ and the JJs have a nominal area of $0.3 \times 0.1$ $\mu$m$^2$ with $I_{\rm{c}} \sim 1.6$~$\mu$A. The SQUID is galvanically connected to two $2.5 \times 1.8$ mm$^2$ antenna pads that are located on either side of the SQUID, each via a $110 \times 2500$ $\mu$m$^2$ line, that position them in the centre of the cavities to which they capacitively couple.

\begin{figure}[!hbt]
\vspace{0.0cm}
\includegraphics[scale=1.0]{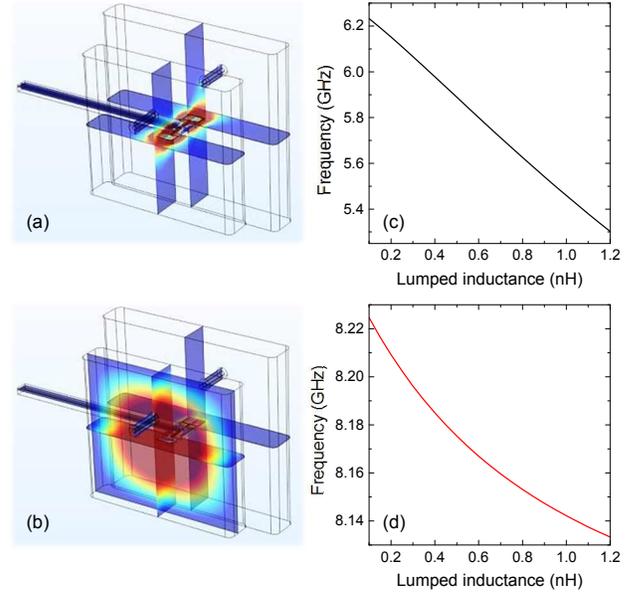}
\vspace{0.0cm}
\small{\caption{(a, b) The dipole-like electromagnetic field distribution of the SQUID resonator and the TE$_{101}$ mode in the HF cavity respectively with the SQUID mimicking lumped element inductance set at 1.2 nH. (c, d) The frequency of the SQUID resonator and the response of the HF cavity's resonance respectively as the SQUID mimicking lumped element inductance is varied.}}
\end{figure}

In order to understand the dynamics of this 3D variant of a JPA, its resonant electromagnetic field distributions are numerically evaluated via the RF module in COMSOL \cite{CS}. Specifically all the electromagnetic resonances were extracted as a function of the SQUID inductance, which is modelled as a lumped element, and the model included both cavities, the SQUID chip, the ceramic chip incorporating the gold-plated copper waveguide and the SMA connectors. The simulations reveal the circuit embedding the SQUID between the large antennas exhibits a resonant electromagnetic mode with a dipole-like field distribution in addition to the cavities sustaining resonant transverse electromagnetic (TE) modes as shown in Figs.~2(a) and 2(b) respectively \cite{SM}. As the SQUID's inductance is varied via the lumped element, from 0.1 nH determined by $I_{\rm{c}}$ at zero magnetic field, the dipole-like resonance in the SQUID resonator at 6.2 GHz is reduced by $\sim1$ GHz as shown in Fig.~2(c), and in response, the HF cavity exhibits a red shift over 90 MHz starting at 8.22 GHz as shown in Fig.~2(d). The apparent linear frequency shift of the former stems from the lumped element inductance variations being much smaller than the numerically evaluated 8.6 nH inductance of this circuit from Fast Henry \cite{FH}. The COMSOL simulation indicates a circuit quantum electrodynamics-like dispersive interaction between the SQUID resonator and the 3D cavity with the latter inheriting a Kerr anharmonicity in its restoring potential that is diluted by the large frequency separation of these resonances \cite{3Djpa1, 3Djpa2, 3Djpa3}. Indeed, the dispersive interaction scales with this frequency separation and the LF cavity exhibits a stronger response as consequence of its spectral proximity to the SQUID resonance \cite{SM}. 

\begin{figure}[!hbt]
\vspace{0.0cm}
\includegraphics[scale=1.0]{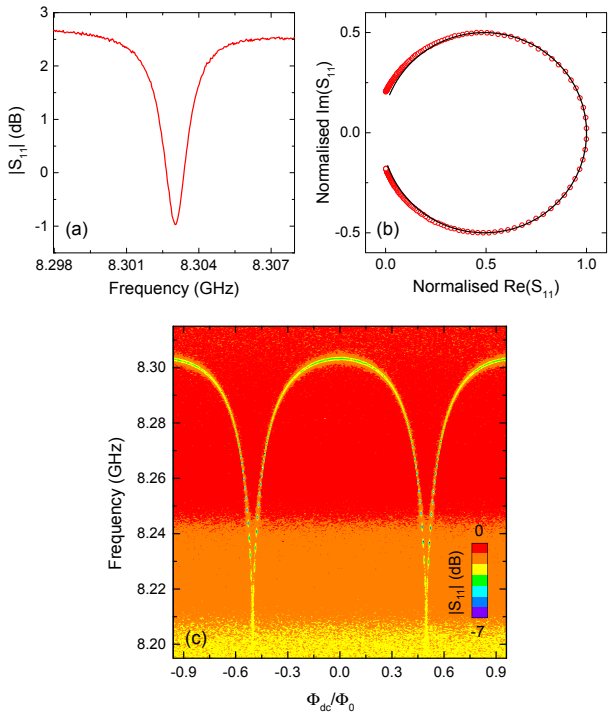}
\vspace{0.0cm}
\small{\caption{(a) The raw reflection coefficient $|S_{11}|$ as a function of frequency for the HF cavity acquired with a VNA using a probe power of $-110$ dBm at the cavity input. (b) The corresponding normalised complex reflection coefficient (points) and the analytical fit (line) used to extract the internal loss and external coupling rates. (c) $|S_{11}|$ as a function of frequency measured via a VNA with a probe power of $-115$~dBm, whilst dc current is applied to the flux line to modulate the SQUID inductance.}}
\end{figure}

Experimentally, in order to modulate the SQUID's inductance, a current line is defined 1 $\mu$m away, as shown in the bottom right of Fig.~1(a), which corresponds to a numerically evaluated mutual inductance of $\approx 5$ pH \cite{3DjpaV1, FH}. This line is connected to a coplanar waveguide with 50~$\Omega$ impedance which is also patterned onto the SQUID chip via electron-beam lithography. This in turn is wire-bonded to a second coplanar waveguide patterned on a temperature stable ceramic (TMM10i, Rogers Corporation), which is located in another tunnel between the cavities. Finally this is connected to an SMP connector which enables external ac and dc currents to be inductively coupled to the SQUID thus enabling it to be externally flux biased and pumped at frequency $\omega_{\rm{p}}$ namely $\Phi_{\rm{ext}}=\Phi_{\rm{dc}}+\Phi_{\rm{ac}}\cos(\omega_{\rm{p}}t)$. Conversely, the cavities are probed via homodyne spectroscopy from SMA connectors directly mounted on the cavities whose external coupling rates were tuned at room temperature by simply adjusting the length of their centre pins.

The experiments were performed in a dry dilution refrigerator, shielded from magnetic noise, with the sample mounted on the base plate at $<10$ mK and measured using a standard microwave setup as detailed in Fig.~1(b). The HF cavity was probed in the spectral region suggested by the COMSOL simulation in Fig.~2(d) via a reflection measurement, using the circuit depicted in green in Fig.~1(b), whilst the flux line was deactivated. The result, shown in Fig.~3(a), reveals a microwave resonance around the anticipated frequency. The resonance linewidth was evaluated by fitting the corresponding normalised complex reflection coefficient from standard microwave circuit theory, as shown in Fig.~3(b), which revealed the HF cavity is over-coupled with an external coupling rate of 1.1 MHz and internal loss rate 0.42 MHz \cite{uWbook}.

To appraise the evolution of the cavity resonance as the capacitively coupled SQUID resonator's inductance is modulated, the previous measurement is repeated but now with the application of dc current bias ($I_{\rm{dc}}$) to the flux line for $\Phi_{\rm{dc}}$, as depicted in magenta in Fig.~1(b). The result shown in Fig.~3(c) reveals the cavity frequency can be periodically modulated as the magnetic flux piercing the SQUID is varied. The experimental nonlinear and periodic magnetic flux dependence of the SQUID resonance cannot be modelled in COMSOL and hence the profile of the HF cavity's dispersive response is also not accurately captured in Fig.~2(d). Nevertheless, at the half magnetic flux quantum $\Phi_0/2$, the cavity frequency shift of $\sim90$ MHz is in-line with the COMSOL simulation, indicating that the SQUID's inductance has a maximal variation of 1.1 nH.

Given the flux dependent frequency modulations in the 3D cavity, the availability of parametric amplification can now be explored where in addition to a weak {\it signal} input at frequency $\omega_{\rm{s}}$, the flux line is simultaneously {\it pumped} to give $\Phi_{\rm{ac}}$ at frequency $\omega_{\rm{p}}=2\omega_{\rm{s}}+\Delta$ whilst $\Phi_{\rm{dc}}$ is varied. This results in signal amplification in the non-degenerate regime when $\Delta \neq 0$ and the concomitant generation of an {\it idler} at  $\omega_{\rm{i}}=\omega_{\rm{s}}+\Delta$ \cite{jpa3, jpa5}. The result of this measurement, shown in Fig. 4(a), reveals that the signal (idler which is not shown) can only be amplified (generated) at specific dc flux biases. This observation arises from the absence of flux dependent frequency modulations at $\Phi_{\rm{dc}}/\Phi_0 \approx$ an integer. On the other hand when $\Phi_{\rm{dc}}/\Phi_0 \approx (2n+1)/2$, where $n=$ integer, the SQUID's critical current becomes so small that it can easily be quenched by the circulating current in the SQUID induced via $\Phi_{\rm{ac}}$.

\begin{figure}[!hbt]
\vspace{-0.0cm}
\includegraphics[scale=1.0]{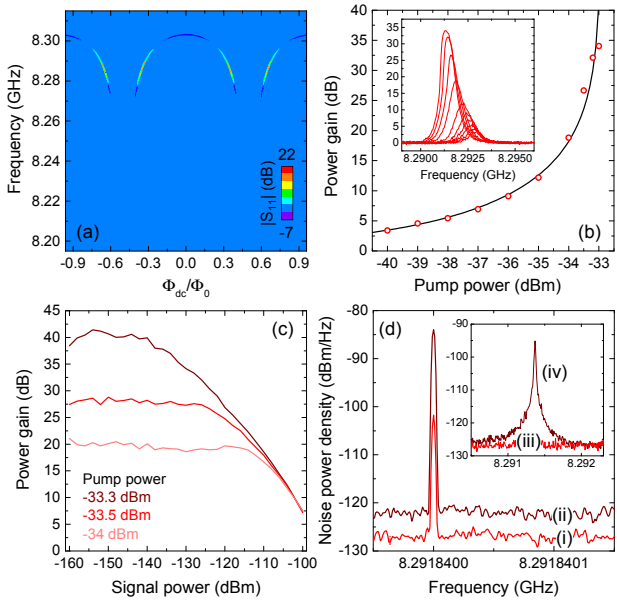}
\vspace{0.0cm}
\small{\caption{(a) $|S_{11}|$ as a function of frequency and dc flux bias with signal power = $-120$ dBm, pump power = $-34$~dBm and $\Delta/2\pi=30$ Hz. (b) The maximum signal power gains (points) extracted from the data shown in the inset, with the cavity measured via signal power = $-130$ dBm, $\Phi_{\rm{dc}}/\Phi_0 = 0.3$ and $\Delta/2\pi = 20$ Hz. Also shown is an analytical fit (line) detailed in the main text. (c) The non-degenerate signal power gain measured at $\Phi_{\rm{dc}}/\Phi_0 = 0.3$ for various ac pump and signal powers. (d) The noise power density measured in a spectrum analyser via a -140~dBm signal tone with: (i) the pump de-activated, (ii) activated at $-34$~dBm. The latter configuration yields $\sim20$ dB parametric amplification and a concomitant signal-to-noise ratio improvement. The inset indicates in the absence of a signal tone: (iii) the cryogenic amplifier's noise when the 3D JPA is de-activated (iv) activating the pump at $-32.8$~dBm results in a parametric resonance.}}
\end{figure}

In order to explore optimal non-degenerate parametric amplification, a dc flux bias is selected between these extremes~[Fig.~4(a)]. In the inset to Fig.~4(b), signal amplification in the HF cavity can be observed  as the ac pump power $P$ is increased. The resultant maximum power gain is plotted in Fig.~4(b) and is limited by the activation of a parametric resonance above a critical pump power $P_{\rm{c}}$, namely when the cavity loss is completely neutralised by the parametric gain, as shown in the inset to Fig.~4(d) \cite{jpa4, jpaq2}. In the limit $\Delta \rightarrow 0$, the signal power gain can be approximated to: $1+ (4P/P_{\rm{c}})/(1-P/P_{\rm{c}})^2 $ which results in a fit, shown in Fig.~4(b), that can reproduce the experimental response and thus confirm the availability of non-degenerate parametric amplification \cite{jpa11}. Finally as the gain increases the cavity linewidth also narrows and at a signal power gain of 20 dB, a 0.4 MHz instantaneous bandwidth can be extracted \cite{jpa5}.

Now to evaluate the enhancement in the dynamic range of the 3D JPA as a linear amplifier, the signal power gain over a large range of input powers is measured with a number of pump powers as summarised in Fig.~4(c). Non-degenerate gains in excess of 40 dB are achieved for signal powers below $-140$ dBm. On the other hand, the minimum gain needed to overwhelm the following commercial cryogenic amplifier noise is typically 20 dB \cite{jpa5} and this gain can be achieved with a 1 dB compression point of approximately $-115$ dBm.

To exploit this performance, the noise added by the 3D JPA needs to be evaluated to confirm if this amplification is available close to the quantum limit. For this purpose the approach of signal-to-noise ratio improvement is employed \cite{jpa3, jpa13}. To that end, the noise power in the spectral region around the cavity resonance is acquired via a spectrum analyser, with the 3D JPA off and on where the latter is set to yield $\sim20$ dB gain, as shown in Fig.~4(d). This measurement reveals a mean noise rise of $4.8 \pm 0.25$~dB with one standard deviation uncertainty and signal gain $G_{\rm{JPA}}=17.8$ dB, which is less than expected due to the signal tone being slightly detuned from the cavity resonance. A $13 \pm 8.5$ dB improvement in the signal-to-noise ratio is observed when the 3D JPA is activated and this can be equated to $1/((T_{\rm{JPA}}/T_{\rm{cryo}})+(1/G_{\rm{JPA}}))$ \cite{jpa9, jpa13}. The cryogenic amplifier's noise temperature, $T_{\rm{cryo}}=4.4 \pm 0.3$ K, is calibrated using the approach detailed in Ref.~\citenum{HEMTnoise} and is summarised in Supplemental Materials \cite{SM}. The 3D JPA's noise temperature $T_{\rm{JPA}}$ is then determined to be $0.148 \pm 0.098$~K and this can be contrasted with vacuum noise $\hbar \omega/2k_B =0.199$ K at 8.3 GHz where $\hbar$ is the reduced Planck constant and $k_B$ is the Boltzmann constant. Consequently, the 3D JPA adds only half a quantum of noise during signal amplification, the minimum possible for a non-degenerate phase-insensitive parametric amplifier  \cite{qlm1, jpa3, jpa4, jpa11}.

To put these performance metrics into context, the highest 1 dB compression point of $-99$ dBm, with a power gain of $\sim 20$ dB in a few GHz bandwidth, was observed in a travelling wave parametric amplifier \cite{jpa8, jpa10}. Although the 3D JPA can offer even greater gains as well as a respectable 1 dB compression point, it suffers in terms of its limited bandwidth. However, the bandwidth available to the 3D JPA can easily be broadened by increasing the external coupling rate by simply adjusting the length of the SMA connector. For parametric amplification to still occur, this enhanced bandwidth needs to be narrower than the dc current induced frequency modulations which suggests that at least 10 MHz should be possible.

On the other hand, operating the 3D JPA in a degenerate phase-sensitive configuration should not only yield gains in excess of 40 dB \cite{jpa5} but it could also achieve unprecedented levels of vacuum squeezing which could have applications for metrology and this is a natural future direction of study \cite{jpaQO1, jpa11}. Finally, the relative ease at which a parametric resonance can be activated in the low-loss 3D cavity, as shown in the inset to Fig.~4(d), paves the way towards higher-order parametric resonances which could yield non-trivial macroscopic quantum states \cite{JPM}. 

A JPA has been executed by capacitively coupling a flux pumped SQUID resonator to a 3D microwave cavity. The dispersive interaction between them results in the 3D cavity's resonance inheriting a diluted Kerr nonlinearity which enables the resultant JPA to yield non-degenerate signal gains exceeding 40 dB. On the other hand, a 20 dB gain can be achieved with half a quantum of added noise and a 1 dB compression point of $-115$ dBm. The 3D JPA also lays the foundations to study quantum optics in massive 3D microwave cavities. 

The authors thank S. Kono and A. Noguchi of the University of Tokyo for their support during the inception of this project. This work was partially supported by CREST(JPMJCR1774) and ERATO (JPMJER1601), JST.


%

\end{document}